\documentclass{article} 
\usepackage{iclr2023_conference_tinypaper,times}

\usepackage{amsmath,amsfonts,bm}

\def\eqref#1{equation~\ref{#1}}

\def\1{\bm{1}}

\DeclareMathAlphabet{\mathsfit}{\encodingdefault}{\sfdefault}{m}{sl}
\SetMathAlphabet{\mathsfit}{bold}{\encodingdefault}{\sfdefault}{bx}{n}

\usepackage{hyperref}
\usepackage{url}
\usepackage{graphicx}

\title{When SMILES have Language: \\ Drug Classification using Text Classification Methods on Drug SMILES Strings}

\iclrfinalcopy

\author{Azmine Toushik Wasi$^1$, Karlo Serbetar$^2$, Raima Islam$^3$, Taki Hasan Rafi$^4$ \& Dong-Kyu Chae$^4$\thanks{Corresponding author.}  \\
$^1$Shahjalal University of Science and Technology, Bangladesh
\texttt{azmine32@student.sust.edu} \\
$^2$University of Cambridge, United Kingdom
\texttt{ks798@cantab.ac.uk}\\
$^3$BRAC University, Bangladesh
\texttt{raima.islam@g.bracu.ac.bd} \\
$^4$Hanyang University, Republic of Korea
\texttt{\{takihr, dongkyu\}@hanyang.ac.kr}
}

\begin{document}

\maketitle

\begin{abstract}
Complex chemical structures, like drugs, are usually defined by SMILES strings as a sequence of molecules and bonds. These SMILES strings are used in different complex machine learning-based drug-related research and representation works. Escaping from complex representation, in this work, we pose a single question: What if we treat drug SMILES as conventional sentences and engage in text classification for drug classification? Our experiments affirm the possibility with very competitive scores. The study explores the notion of viewing each atom and bond as sentence components, employing basic NLP methods to categorize drug types, proving that complex problems can also be solved with simpler perspectives. The data and code are available here: \url{https://github.com/azminewasi/Drug-Classification-NLP}.
\end{abstract}

\section{Introduction}
Classifying drug types plays a pivotal role in drug discovery research, aiding in the categorization of established drugs and enhancing our understanding of the distinctive features of newly identified or synthesized drugs. 
It is necessary to ensure that a drug is used safely and that you get the greatest possible benefit with the lowest possible risk.
Different deep generative models have demonstrated efficacy in addressing various drug discovery challenges \citep{pandey2022transformational}, mostly with the capabilities of utilizing complex chemical structural data.

Simplified Molecular Input Line Entry System (SMILES) is a text-based representation of a chemical molecule \citep{10.3389/fphar.2022.1046524}. They provide a standardized language for encoding molecular information, facilitating analysis and machine learning applications in drug-related research. One example of a drug structure and corresponding SMILES is shown in figure \ref{fig:example}.

In this study, we explore the drug classification challenge from a simple perspective using drug SMIILES. Given that drug chemical structures are conventionally denoted through SMILES strings, an opportunity arises to avoid complex chemical representations by considering drug SMILES as simple text sentences. In this analogy, the individual atoms and bonds within the molecule serve as the constituent words, forming a coherent sentence using the sequential arrangement of SMILES, word after word. Experimental results show that applying a basic bag-of-n-grams model can achieve very competitive scores, showing proof that simple NLP approaches can be applied to complex problems too, without using any complex chemical embedding or pre-trained models.

\begin{figure}[t]
\begin{center}
\includegraphics[scale=0.5]{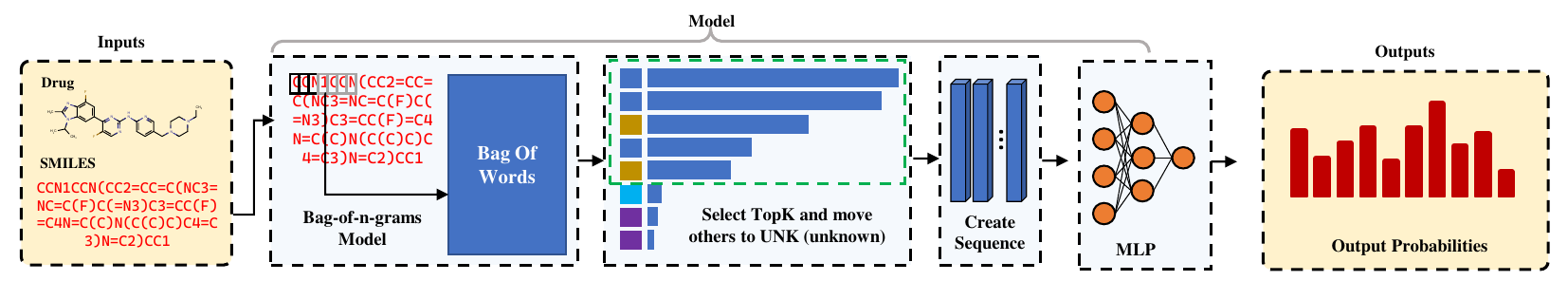}
\end{center}
\caption{Overview of our approach for Drug Classification using Text Classification Methods on Drug SMILES Strings }\label{fig:model}
\end{figure}

\section{Method}
In the process, a given SMILES string undergoes encoding via the bag-of-n-grams model, where `n' signifies the number of letters within each token. Multiple tokens are generated from the dataset, with the top `K' (the most frequently occurring) tokens selected. The remaining tokens are amalgamated into an `unknown' token (`UNK'). Subsequently, a sequence is constructed based on the frequency of token appearances. This string is then fed into a Multilayer Perceptron (MLP) to obtain logits, from which a classification label is determined by selecting the one with the highest probability. A visual overview of the approach is presented in figure \ref{fig:model}. 

\section{Experiment}
In our experimental setup, we utilized a drug dataset obtained from \cite{dataset}, partitioned into 70\% for training, 10\% for development, and 20\% for testing. The dataset has 12 classes; they are: dermatologic, antiinfective, antineoplastic, CNS, hematologic, lipidregulating, antiinflammatory, cardio, gastrointestinal, respiratory system, reproductive control, and urological. More dataset information is included in Appendix \ref{dataset-apps}. The modeling approach involved considering combinations of 1 to 5 grams, representing sequences of 1 to 5 letters in the tokens, while encoding each model. Each experiment is performed multiple times with different seeds, and an average is taken. {Also, we have performed fine-tuning only on $TopK$, and the best value is $1250$. Experimental setup details with comparison and ablation studies are presented in Appendix \ref{E-Appendix}. Here in table \ref{tab:exp-table}, we present comparative analysis between different methods For fingerprints, we use these fingerprints: Atom Pair Fingerprint, MACCS Fingerprint, Morgan Fingerprint.}

\begin{table}[h]
\centering
\small
\caption{Performance Metrics for Different Configurations}
\label{tab:exp-table}
\begin{tabular}{l c c c c c c }
\textbf{Model} & \textbf{Accuracy} & \textbf{Precision} & \textbf{Recall} & \textbf{F1 (Weighted)} & \textbf{F1 (Macro)} & \textbf{ROC-AUC} \\
\hline
1-gram+MLP &  0.622 & 0.610 &  0.622 &  0.604 &  0.406 & 0.760 \\
2-gram+MLP  &  0.669 & 0.700 &  0.669 &  0.672 &  0.445 & 0.810 \\
3-gram+MLP  & \bf 0.737 & \bf 0.764 &  \bf 0.737 & \bf 0.744 &  0.553 & \bf 0.848  \\
4-gram+MLP & 0.726 & 0.758 &  0.726 &  0.731 &  0.524 & 0.841  \\
5-gram+MLP  & 0.728 & 0.740 &  0.728 &  0.730 &  \bf 0.563 & 0.838  \\
\hline
AtomPair+MLP& 0.799& 0.804&  0.800&  0.799&  0.702& 0.876\\
MACCS+MLP& 0.797& 0.801&  0.797&  0.796&  0.702& 0.873\\
Morgan+MLP& \bf 0.800& \bf 0.804&  \bf 0.800&  \bf 0.799& \bf 0.703& \bf 0.876
\end{tabular}
\end{table}

{
Table \ref{tab:exp-table} illustrates the performance of several drug classification models. Among the ngram models, 3-gram models achieve around 73.7\% accuracy and 76.4\% precision in our experimental setup. Most of the ROC-AUC scores are also above 0.835, suggesting good performance. Molecular fingerprint models like AtomPair+MLP, MACCS+MLP, and Morgan+MLP exhibit improved accuracy and noteworthy precision, recall, and F1 scores. Remarkably, Morgan+MLP excels in various metrics, showcasing its effectiveness despite the advantage of molecular fingerprints. In summary, 3-gram+MLP emerges as the optimal solution among ngrams, showcasing competitive scores with fingerprint-based models. This demonstrates the feasibility of treating drug SMILES as strings and employing basic NLP methods for classification tasks. While molecular fingerprints are intended to capture particular molecular characteristics and are therefore anticipated to score higher, it is noteworthy that n-gram models, such as 3-gram+MLP, surprisingly hold their own and do reasonably well in drug classification.
This observation emphasizes the versatility and competitive performance of n-gram approaches, even when compared to specialized molecular fingerprinting techniques.}
Additional details on our model's correlation with fingerprint-based models, our work's practical significance, and its limitations are discussed in Appendix \ref{app-disc}.




\subsubsection*{Acknowledgements}
We appreciate the reviewers' insightful feedback, which helped us to improve the paper's quality. The work of Dong-Kyu Chae and Taki Hasan Rafi was supported by (1) the National Research Foundation of Korea (NRF) grant funded by the Korea government (*MSIT) (No.2018R1A5A7059549) and (2) the Institute of Information \& communications Technology Planning \& Evaluation (IITP) grant funded by the Korea government (MSIT) (No.2020-0-01373,Artificial Intelligence Graduate School Program (Hanyang University)). *Ministry of Science and ICT

\subsubsection*{URM Statement}

Authors Azmine Toushik Wasi and Raima Islam meet the URM criteria of the ICLR 2024 Tiny Papers Track.

\bibliography{iclr2023_conference_tinypaper}
\bibliographystyle{iclr2023_conference_tinypaper}

\appendix
\section{Related Works} \label{RW}
\textbf{Deep learning and NLP:} Deep learning (DL) models are utilized in drug development for the following purposes: quantitative structure-activity relationship, virtual screening, and drug design \citep{1-10.1093/bib/bbab430}. Therefore, in recent years, we have seen the application of various DL systems being employed for drug tasks where \citep{2-xiong2019pushing} incorporated graph attention methods to Graph Neural Networks and constructed Attentive FP, a function capable of preserving the interactions between topologically adjacent atoms. \citep{3-gomez2018automatic} devised a Variational Autoencoder model to facilitate the automated design of molecules to transform the SMILES input strings into a representation of continuous vectors. Based on SeqGAN \citep{5-yu2017seqgan}, \citep{4-guimaraes2017objective} constructed objective-reinforced Graph Adversarial Network (ORGAN) to produce molecules from SMILES sequences while optimizing a variety of domain-specific metrics. 
For text classification, n-gram modeling is one of the most basic and fundamental model. \citep{6-khan2022multi} have employed several n-gram features, including unigrams, bigrams, trigrams, and numerous combinations thereof to train DL classifiers and perform sentiment analysis, achieving a good score. In our work, we have also presented an example of how effective it can be in modeling very complex scenarios like drug classification.  

{
\textbf{SMILES-drug representation:} While recent deep learning advances have resulted in accelerating drug discovery processes, not many of them address the generalization problem due to lack of labeled data. \citep{wu2021learning-11} develop a bidirectional long-short-term memory attention network (BAN)  with a multi-step attention framework, extracting the important characteristics  from SMILES strings and capturing latent representations of molecules. Tackling the same problem of data availability, \citep{winter2022smile-12} leverage an NLP mechanism called SMILES-to-properties-transformer (SPT) for predicting binary limiting activity coefficients from SMILES codes for thermodynamic property prediction. Through the integration of synthetic and experimental data, the fine-tuning process achieved computational efficiency.} 

{
\textbf{Fingerprint-based representation:} \citet{ali2023when} presented a detailed study of fingerprint-based feature extraction for drug subcategory classification. This work explored the prediction of drug subcategories by employing traditional molecular fingerprints and sequence-based embedding methods, specifically focusing on SMILES strings in the bioinformatics domain. The study evaluates five types of embeddings, including Morgan fingerprint, MACCS fingerprint, k-mers, and minimizer-based spectrum. Furthermore, a weighted variant of k-mers, incorporating inverse document frequency for assigning weights to individual k-mers within the spectrum, is also investigated.}

{\textbf{Graph-based Drug representation:} Graph ML has established its usefulness in biomedical applications, especially in classification tasks. \citet{NIPS2019_9054} developed N-Gram Graph, an unsupervised molecular representation. By embedding vertices in the molecule graph and constructing compact representations through short walks, this method successfully represents molecular properties; empirical experiments and theoretical analyses confirm the method's strong representation and prediction capabilities. In contrast, our model adopts a simpler perspective, leveraging drug SMILES as text sentences for drug classification directly without any complexities.}


\section{Experiments} \label{E-Appendix}
\subsection{Dataset} \label{dataset-apps}
{
In our study, we utilized a drug dataset retrieved from \cite{dataset}, partitioning it into training (70\%), development (10\%), and testing (20\%) subsets for analysis. We removed all multi-label options and kept only single labels for this work to simplify the model. The dataset is available in the supplementary materials, and the distribution of labels across the 12 classes is outlined in Table \ref{tab:dataset-classes}. "Antiinfective" drugs are most common, followed by "antineoplastic" and "CNS". There is a class-imbalance between different classes. The most common type "antiinfective" is 83 times more present than the least common type "urological".}

\begin{table}[h]
\centering
\caption{Dataset Classes}
\label{tab:dataset-classes}
\begin{tabular}{lc}
\hline
\textbf{Type} & \textbf{Count} \\
\hline
antiinfective & 2412 \\
antineoplastic & 1175 \\
cns & 1149 \\
cardio & 797 \\
antiinflammatory & 372 \\
hematologic & 266 \\
\end{tabular}
\begin{tabular}{lc}
\hline
\textbf{Type} & \textbf{Count} \\
\hline
gastrointestinal & 259 \\
lipidregulating & 164 \\
reproductivecontrol & 148 \\
dermatologic & 115 \\
respiratorysystem & 100 \\
urological & 29 \\
\end{tabular}
\end{table}

\begin{figure}[h]
\begin{center}
\vspace{-4mm}
\includegraphics[scale=0.2]{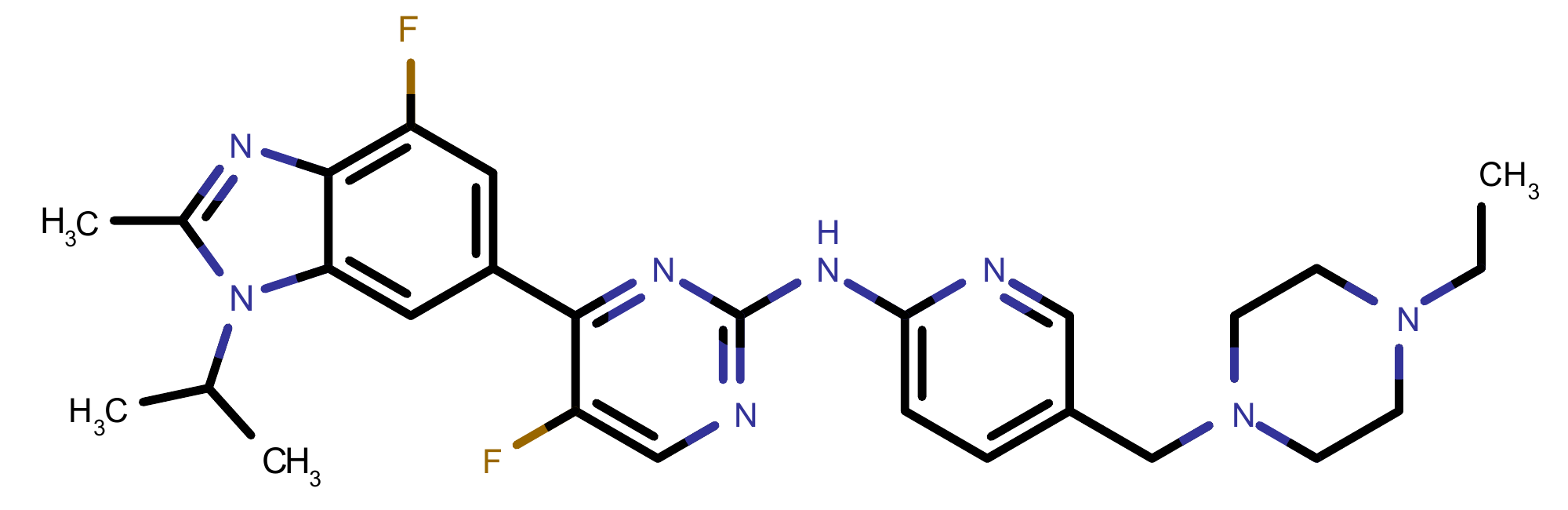}
\end{center}
\caption{Molecular structure for a drug named \textbf{Abemaciclib}, with following SMILES string - \texttt{CCN1CCN(CC2=CC=C(NC3=NC=C(F)C(=N3)C3=CC(F)=C4N=C(C)N(C(C)C)C4=C3)N=C2)CC1} \citep{drug-bank}. }\label{fig:example}
\vspace{-1mm}
\end{figure}

{
Here we present the n-gram modeling of a drug molecule SMILES, named \textbf{Abemaciclib}. The SMILES string is:
\texttt{CCN1CCN(CC2=CC=C(NC3=NC=C(F)C(=N3)C3=CC(F)=C4 N=C(C)N(C(C)C)C4=C3)N=C2)CC1}.  Table \ref{tab:exp-totalK} shows the number of unique n-grams for each n, which is denoted by $K$ for this particular SMILES. }

\begin{table}[h]
\centering
\caption{n and K in \textbf{Abemaciclib} }
\label{tab:exp-totalK}
\vspace{2mm}
\begin{tabular}{ccc}
\textbf{n} & \textbf{K} & \textbf{Some Examples} \\
\hline
1 & 10 & `N', `C', `F', `='\\
2 & 31 & `NC', `C2', `C3', `C=', `CN', `N=' \\
3 & 58 &  `C4N=', `4=C3', `=C4', `N1C'\\
4 & 67 & `C3=C',  `CCN1', `C3=N', `C4N=', `=NC='  \\
5 & 69 & `(=N3)', `CC2=C', `C(F)C' \\
\end{tabular}
\end{table}

\subsection{Experimental Details}  \label{ED-Appendix}
Table \ref{tab:experiment_parameters} presents experimental parameters and hyperparameters. In all experiments, the parameters are kept the same. Experimental codes are included in supplementary materials.

\begin{table}[h]
\centering
\caption{Parameter Configuration for the Experiment}
\label{tab:experiment_parameters}
\begin{tabular}{l c}
\bf NAME & \bf VALUE \\
\hline
Number of Epochs & 200 \\
TopK & 1250 \\
Dev Set Ratio & 0.1 \\
Test Set Ratio & 0.2 \\
Number of ngrams considered (TopK+1) & 1251 \\
Batch Size & 32 \\
MLP Hidden Size & [512, 256, 128, 32]\\
MLP Dropout & 0.1 \\
MLP Learning Rate & $3 \times 10^{-5}$ \\
\end{tabular}
\end{table}

\subsection{Ablation Study: TopK} \label{ED-Ablation}
{In Table \ref{tab:topk_table}, we present an ablation study on the hyperparameter topK using 3-grams. The results indicate optimal performance at $TopK = 1250$, followed closely by $TopK = 1500$ as the second-best configuration.}

\begin{table}[h]
\centering
\caption{Performance at Different Values of TopK}
\label{tab:topk_table}
\begin{tabular}{l c c c c c c c}
\textbf{TopK} & \textbf{Accuracy} & \textbf{Precision} & \textbf{Recall} & \textbf{F1 (W.)} & \textbf{F1 (Macro)} & \textbf{ROC-AUC} & \textbf{AUPRC} \\
\hline
500 & 0.705 & 0.744 &  0.705 &  0.713 &  0.525 & 0.830 &  0.587 \\
750 & 0.695 & 0.715 &  0.695 &  0.697 &  0.435 & 0.825 &  0.574 \\
1000  & 0.711 & 0.740 & 0.711 & 0.717 & 0.477 & 0.834 & 0.593 \\
1250 & \bf 0.737 & \bf 0.764 &  \bf 0.737 & \bf 0.744 &  0.553 & \bf 0.848 &  \bf 0.620 \\
1500 & 0.732 & 0.758 &  0.732 &  0.737 &  \bf 0.561 & 0.846 &  0.612 \\
\end{tabular}
\end{table}

\section{Discussion} \label{app-disc}
\subsection{Discussion on Experimental Findings}
{Drug classification is challenging, considering molecular structures are complex and multifaceted. The task is further complicated by the broad spectrum of compounds, minor variations in chemical composition, and the constantly changing interactions that occur within biological systems. Furthermore, the dynamic landscape of drug design and discovery demands flexible classification models that are capable of navigating novel chemical spaces, which makes the task inherently demanding}. 

{ Molecular fingerprints and n-gram modeling serve as indispensable tools in Cheminformatics and language modeling, respectively, enabling the efficient comparison and analysis of complex structures. In order to extract features, molecular fingerprints transform structural elements into bits in a bit vector or counts in a count vector, such as substructures for small molecules or atom-pairs for larger molecules \cite{Riniker2013}. Morgan fingerprints, a type of molecular fingerprint, transform intricate molecular structures into unique bit vectors by identifying circular substructures within a specified radius around each atom \cite{Capecchi2020}. On the other hand, n-gram modeling operates as a probabilistic language model, predicting the likelihood of word sequences by breaking down input into chunks (n-grams) and assessing their probabilities based on occurrence frequency. The goal of both methods is to reduce complex structures—like molecules or strings—to simpler, more similar forms. The correlation is also observed in our model's scores, as presented in Tables \ref{tab:exp-table} and \ref{tab:topk_table}.
This interesting phenomenon suggests a relationship between the chemical aspects of our model and the corresponding parameters in Morgan fingerprints. For instance, the optimal performance observed at $TopK=1250$ and $n=3$ mirrors the characteristics of Morgan fingerprints, where $TopK$ is associated with the standard number of bits and $n$ with the specified radius. The close alignment of our model's optimal settings with Morgan fingerprints exemplifies how chemical intuition can fine-tune simple n-gram models in Cheminformatics intuitively. Further research could delve into these rationales and explore these intriguing relationships. }

{ Apart from these shared characteristics, there are also some differences. Morgan fingerprints are susceptible to bit collisions when different substructures map to the same bit, which could lead to ambiguity. On the contrary, n-gram models perform very well in representing each n-gram distinctively, which neutralizes the risk of collisions. Furthermore, molecular fingerprints only account for the presence of specific substructures, potentially leaving out vital information in some cases, whereas n-gram models are context-sensitive and take letter order of SMILES into consideration, which can induce noise.}

{Our work also suggests that capturing some key substructures is sufficient for drug classification. Table \ref{tab:topk_table} shows that the score increases until $TopK$ reaches 1250 and declines thereafter, indicating that only 1250 tokens are sufficient for accurate classification. This proves that, by focusing on essential molecular motifs, models are able to distill meaningful information, striking a balance between computational efficiency and predictive accuracy. This approach acknowledges the practical constraints of analyzing large chemical spaces while still yielding valuable insights for effective drug classification by proposing a simple baseline.}

\subsection{Discussion on Practical Impact and Scalability}
{We believe that using a basic NLP model in drug classification has significant potential impact and utility. This approach streamlines the representation of complex chemical structures into a format analogous to natural language, thereby simplifying the drug classification process. The model's success in achieving competitive scores without relying on intricate chemical embeddings or pre-trained models demonstrates its effectiveness in addressing complexity through simplicity.
This technique has the potential to transform drug classification by providing a more accessible and interpretable framework, potentially enhancing collaboration between experts in diverse fields. Its simplicity not only promotes ease of implementation but also contributes to democratizing drug discovery processes, making them more approachable for researchers and practitioners without extensive expertise in chemoinformatics.}

\subsection{Discussion on Limitations and Future Works}
{In addition to drug classification, NLP-based methods can play a pivotal role in drug Quantitative Structure-Activity Relationship (QSAR) research by enabling the extraction of meaningful information from textual data, enhancing the understanding of drug properties and interactions through advanced linguistic analysis. 
Also, we can observe the class imbalance in the table \ref{dataset-apps}. Future works in this area may explore strategies like oversampling, undersampling, or synthetic data generation to address this issue. Additionally, leveraging advanced transfer learning models may enhance adaptability, presenting promising avenues for further investigation into robust drug classification methods using NLP techniques.
The interpretability of this model can be utilized to clarify the decision-making process involved in drug classification, assisting researchers and healthcare providers in comprehending the variables affecting forecasts. The transparency of the model makes it easier to identify the critical characteristics that contribute to each drug type, offering insightful information about the classification choices made.}

\section{Conclusion}
This study showcases the application of fundamental NLP models to intricate challenges like drug classification by treating drug SMILES as strings. Our experimental findings reveal that our basic NLP model, typically defined by a bag-of-n-grams approach, attains highly competitive scores in drug classification tasks.

\end{document}